\documentclass[aps,prc,twocolumn,groupedaddress,nofootinbib]{revtex4-1}

\usepackage{graphicx}

\begin{document}


\title{Shape evolution of Zr nuclei and roles of tensor force}


\author{S. Miyahara}
\author{H. Nakada}
\email[E-mail:\,\,]{nakada@faculty.chiba-u.jp}

\affiliation{Department of Physics, Graduate School of Science,
 Chiba University,\\
Yayoi-cho 1-33, Inage, Chiba 263-8522, Japan}


\date{\today}

\begin{abstract}
  Shape evolution of Zr nuclei are investigated
  by the axial Hartree-Fock (HF) calculations
  using the semi-realistic interaction M3Y-P6,
  with focusing on roles of the tensor force.
  Deformation at $N\approx 40$ is reproduced,
  which has not been easy to describe
  within the self-consistent mean-field calculations.
  The spherical shape is obtained in $46\leq N\leq 56$,
  and the prolate deformation is predicted in $58\leq N\leq 72$,
  while the shape switches to oblate at $N=74$.
  The sphericity returns at $N=80$ and $82$.
  The deformation in $60\lesssim N\lesssim 70$ resolves the discrepancy
  in the previous magic-number prediction
  based on the spherical mean-field calculations
  [Prog. Theor. Exp. Phys. \textbf{2014}, 033D02].
  It is found that the deformation at $N\approx 40$ takes place
  owing to the tensor force with a good balance.
  The tensor-force effects significantly depend on the configurations,
  and are pointed out to be conspicuous
  when the unique-parity orbit (\textit{e.g.} $n0h_{11/2}$)
  is present near the Fermi energy, delaying deformation.
  These effects are crucial for the magicity at $N=56$
  and for the predicted shape change at $N=74$ and $80$.
\end{abstract}


\maketitle



\section{Introduction\label{sec:intro}}

Shell structure,
which is an obvious quantum effect and is manifested by the magic numbers,
is one of the fundamental concepts
in the nuclear structure physics~\cite{ref:BM1}.
While the spin-orbit ($\ell s$) splitting of the single-particle (s.p.) orbits
is significant in medium- to heavy-mass nuclei,
forming the $jj$-closed magic numbers ($Z,N=28,50,82$ and $N=126$),
it is less important in light nuclei,
where the $\ell s$-closed magic numbers ($Z,N=2,8,20$) are kept.
As the $\ell s$-closed magicity is partly maintained but not so stiffly,
the structure of the Zr (\textit{i.e.} $Z=40$) isotopes strongly depends
on the neutron number $N$,
providing us with a good testing ground of nuclear structure theories.
While the doubly-magic nature of $^{90}$Zr is well known,
it has been established experimentally that Zr nuclei become deformed
both in neutron-deficient (\textit{e.g.} $^{80}$Zr)~\cite{ref:Lis87}
and neutron-rich (\textit{e.g.} $^{100}$Zr) regions~\cite{ref:Che70,ref:NuDat}.
This $N$-dependence is contrasted to the Sn and Pb nuclei,
in which the $Z=50$ and $82$ magic numbers are rigid in a wide range of $N$.
Moreover, the sudden change of the shape from $^{98}$Zr to $^{100}$Zr
was interpreted as a quantum phase transition~\cite{ref:QPT}.
Shape with the tetrahedral symmetry was theoretically argued
for $^{80}$Zr, $^{96}$Zr
and $^{108,110,112}$Zr~\cite{ref:tetra-Zr96,ref:tetra-Zr110,ref:ZLZZ17},
although no experimental evidence has been reported so far.
On the contrary,
the measured first excitation energy at $^{96}$Zr is significantly higher
than in surrounding nuclei~\cite{ref:NuDat,ref:Kha75},
implying submagic nature of $N=56$
as well as the magicity of $Z=40$~\cite{ref:NS14}.
With experiments using radioactive beams
and progress on our understanding of the nuclear shell structure,
it is of interest to reinvestigate ground-state (g.s.) properties
of the Zr nuclei systematically.

One of the recent topics in nuclear structure physics
is roles of the tensor force in the $Z$- or $N$-dependence
of the shell structure
(\textit{i.e.} the shell evolution)~\cite{ref:Vtn,ref:SC14}.
Although the tensor force is contained in the nucleonic interaction,
most of the self-consistent mean-field (MF) calculations
have been performed without the tensor force.
We now face a new problem how the tensor force is incorporated
in the MF framework,
and how it alters pictures obtained in the conventional calculations.
As an attempt to fix this problem,
one of the authors (H.N.) developed
M3Y-type semi-realistic interactions~\cite{ref:Nak03,ref:Nak13}.
In Ref.~\cite{ref:NS14}, a map of magic numbers was drawn
based on the self-consistent MF calculations assuming the spherical symmetry,
by adopting the pairing as a measure of correlations.
It was shown that the semi-realistic interaction M3Y-P6~\cite{ref:Nak13}
gives a prediction of magic numbers compatible with almost all available data,
except $N=20$ in $^{32}$Mg and $Z=40$ in neutron-rich Zr nuclei.
The M3Y-P6 interaction contains the realistic tensor force
originating from the $G$-matrix~\cite{ref:M3Y-P},
which is profitable in reproducing the shell evolution
in some regions~\cite{ref:NS14,ref:NSM13}.
By a recent study, it has been suggested
that the contradiction in $^{32}$Mg mentioned above could be resolved
if the quadrupole deformation is taken into account explicitly~\cite{ref:SNM16}.
It is noted that the deformation in $^{32}$Mg
was obtained via correlations beyond MF,
in the calculations using the Gogny-D1S interaction~\cite{ref:RER00},
although the tensor force has certain effects
on the shape evolution around $N=20$.
It is desired to apply deformed MF calculations also to the Zr nuclei,
and to reexamine the $Z=40$ magicity with the semi-realistic interaction
including the realistic tensor force.

\section{Prediction of deformation at ground state\label{sec:gs-def}}

We have implemented self-consistent HF calculations
assuming the axial symmetry
for the even-$N$ Zr isotopes from $^{80}$Zr to $^{122}$Zr,
and investigate shape evolution for increasing $N$.
The numerical method is detailed in Ref.~\cite{ref:Nak08}.
Since we apply basis functions having good orbital angular momentum $\ell$,
the s.p. space is truncated via the cut-off value $\ell_\mathrm{max}$.
To describe normally-deformed s.p. levels with good precision,
$\ell_\mathrm{max}$ should be greater by four
than the $\ell$ value of the corresponding level
at the spherical limit~\cite{ref:Nak08}.
We therefore adopt $\ell_\mathrm{max}=9$
because of the presence of the $0h_{11/2}$ orbit in $N\leq 82$.
The semi-realistic interaction M3Y-P6~\cite{ref:Nak13} is mainly employed.
For comparison, we have also implemented the axial HF calculations
using the Gogny-D1M interaction~\cite{ref:D1M},
which is one of the most successful interaction so far
but does not include tensor force.
Whereas the D1M interaction was developed for calculations
in which the pairing and the additional quadrupole collective degrees of freedom
(d.o.f.) are taken into account,
it is of interest to compare the results within the HF,
to examine whether or not the tensor-force effects can be imitated
by the other channels at this level.

The minimum giving the lowest energy yields a good candidate
of the g.s. for each nucleus.
However, we ignore pair correlations, rotational correlations,
and triaxial or odd-parity deformation in the present work.
If other minima have close energies to the lowest one,
caution is needed because the ignored correlations
may mix or invert the energies.
Keeping this point in mind,
we shall look at the predicted g.s. deformation of the Zr isotopes.

\begin{figure}
\includegraphics[scale=0.7]{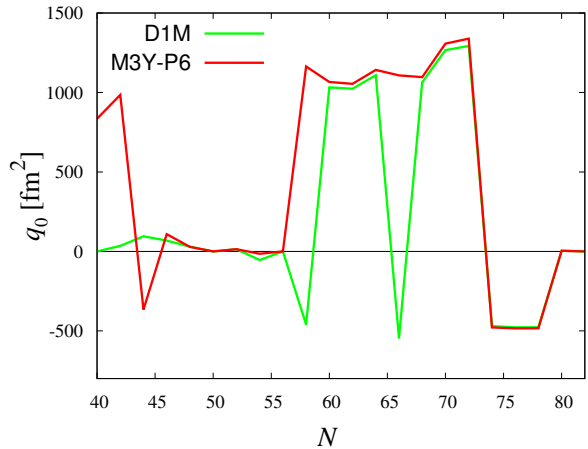}
\caption{(Color online) Values of $q_0$ that give the lowest energy
  for the individual nucleus in the axial HF calculations
  with M3Y-P6 (red line) and D1M (green line).
\label{fig:q0-N}}
\end{figure}

In Fig.~\ref{fig:q0-N}
the $q_0$ values that give the lowest energy at individual $N$ are depicted,
which are obtained from the axial HF calculations with M3Y-P6 and D1M.
Note that $q_0\approx 1000\,\mathrm{fm}^2$
corresponds to the deformation parameter $\beta\approx 0.4$ at $A\approx 100$,
if we apply $q_0 = 3 A\,(1.2A^{1/3})^2 \beta/\sqrt{5\pi}$.

The shape of Zr is indicated to be spherical around $N=50$
in the present calculation, as expected.
It is found that the M3Y-P6 interaction predicts prolate shape
at $N=40$ and $42$,
which seems consistent with the experimental data.
On the contrary, D1M gives spherical shape at $N=40$ at the HF level.
At $N=58$, the absolute minimum is prolate with M3Y-P6,
while oblate with D1M.
In $60\leq N\leq 82$, M3Y-P6 and D1M provide similar deformation
except at $N=66$.
In the M3Y-P6 results,
well-deformed prolate shape gives the lowest energy
for individual nucleus in $58\leq N\leq 72$.
Recall that measured $E_x(2^+_1)$'s are low and close to one another
in $60\leq N\leq 70$~\cite{ref:Sum11,ref:Pau17},
suggesting that these nuclei are well-deformed to a similar degree.
The lowest minimum is switched to the oblate side in $74\leq N\leq 78$,
and returns to the spherical shape at $N=80$.
In $64\leq N\leq 74$,
local minima are observed both on the prolate and the oblate sides,
one of which gives the lowest energy while the other the second lowest,
both in the M3Y-P6 and D1M results.
Whereas the lowest minimum lies on the oblate side at $N=66$
in the D1M result,
the energy difference between the prolate and oblate minima is small
($\approx 0.4\,\mathrm{MeV}$),
preventing us from being conclusive
until taking account of correlations beyond HF.

In Table~\ref{tab:pred-shape}, predicted shapes are compared
among several self-consistent MF calculations.
These results illustrate that shape evolution of Zr can provide
a good testing ground of theoretical approaches,
which are composed of frameworks, methods and inputs.
The results other than the present work are taken from literature.
For the HF results, states with $|q_0|<200\,\mathrm{fm}^2$
are identified to have the spherical shape.
Though not listed in Table~\ref{tab:pred-shape} to avoid overlaps,
Hartree-Fock-Bogolyubov (HFB) results with D1S for $^{98-108}$Zr
were presented in Ref.~\cite{ref:RSRP10},
where the triaxial deformation was taken into account,
and relativistic Hartree-Bogolyubov (RHB) results with DD-PC1 for $^{100-114}$Zr
in Ref.~\cite{ref:ZLZZ17},
where octupole and tetrahedral deformations were also considered.
The HFB results with SLy4 were depicted in Ref.~\cite{ref:DNS02}.
In Ref.~\cite{ref:RSRP10} prolate shape was predicted in $^{100,102,104}$Zr.
No triaxial ground state comes out with D1S, contrary to the RHB results.
All calculations predict deformation in $N\gtrsim 60$,
and most of them give minima on both of the prolate and the oblate sides,
suggesting shape coexistence at low energy.
It depends on the interaction
which of the prolate and the oblate minima is the lowest.

\begin{table}
\caption{Brief summary of shapes of Zr nuclei predicted
  by self-consistent MF calculations;
  spherical (``sph''), prolate (``pro''), oblate (``obl'')
  or triaxial (``tri'').
  For the mean-field type,
  ``(ax.)'' indicates that the axial symmetry is assumed.
  In the row of ``EDF'',
  effective interaction or energy-density functional is specified.
  For the row of ``Ref.'', ``PW'' stands for the present work.
\label{tab:pred-shape}}
\begin{ruledtabular}
\begin{tabular}{cccccccc}
MF & HF (ax.) & HF (ax.) & HFB & HF & HFB (ax.) & RHB & RHB \\
EDF & M3Y-P6 & D1M & D1S & SkM$^\ast$ & SLy4 & NL3$^\ast$ & DD-PC1 \\
Ref. & PW & PW & \protect\cite{ref:D1S-Web} & \protect\cite{ref:INY09}
& \protect\cite{ref:BOUS05} & \protect\cite{ref:AA17} & \protect\cite{ref:AA17}
\\
\hline
$N=40$ & pro & sph & sph & pro & -- & -- & -- \\
$N=42$ & pro & sph & sph & tri & -- & -- & -- \\
$N=44$ & obl & sph & sph & tri & -- & -- & -- \\
$N=46$ & sph & sph & sph & pro & -- & -- & -- \\
$N=48$ & sph & sph & sph & sph & -- & sph & sph \\
$N=50$ & sph & sph & sph & sph & -- & sph & sph \\
$N=52$ & sph & sph & sph & sph & -- & sph & sph \\
$N=54$ & sph & sph & sph & sph & -- & tri & tri \\
$N=56$ & sph & sph & sph & sph & -- & tri & tri \\
$N=58$ & pro & obl & obl & sph & -- & obl & pro \\
$N=60$ & pro & pro & obl & pro & -- & obl & pro \\
$N=62$ & pro & pro & obl & pro & pro & obl & tri \\
$N=64$ & pro & pro & obl & pro & pro & obl & tri \\
$N=66$ & pro & obl & obl & pro & pro & obl & obl \\
$N=68$ & pro & pro & obl & pro & pro & sph & obl \\
$N=70$ & pro & pro & obl & pro & pro & sph & obl \\
$N=72$ & pro & pro & sph & pro & pro & -- & -- \\
$N=74$ & obl & obl & sph & pro & sph & -- & -- \\
$N=76$ & obl & obl & sph & sph & sph & -- & -- \\
$N=78$ & obl & obl & sph & sph & sph & -- & -- \\
$N=80$ & sph & sph & sph & sph & sph & -- & -- \\
$N=82$ & sph & sph & sph & sph & sph & -- & -- \\
\end{tabular}
\end{ruledtabular}
\end{table}

Correlations beyond MF may invert or mix the states with different shapes.
It is sometimes inadequate to conclude from the MF results
what shape the g.s. of each nucleus has.
There have been calculations based on the generator coordinate method (GCM),
by which correlations beyond MF are taken into account~\cite{ref:BHR03}.
However, it should be noticed
that consistency with respect to the effective interactions is lost
in most of the beyond-MF calculations so far,
as the interactions tuned at the MF level are applied.
In this respect, comparison at the MF level is basic
and retains particular significance.
Moreover, density-dependent repulsion,
which is contained in almost all effective interactions
designed for the self-consistent calculations,
gives rise to a serious problem in the approaches
based on the GCM including the symmetry restoration~\protect\cite{ref:Rob10}.

In the subsequent section,
we shall investigate roles of the tensor force in the deformation
for selected nuclei,
via detailed analyses in terms of the energy curves and the s.p. levels.

\section{Energy curves and tensor-force effects\label{sec:energy}}

In this section, energy curves are depicted for several nuclei.
The HF calculations yield local energy minima
depending on the intrinsic mass quadrupole moment $q_0$.
The $q_0$ values at the minima are essentially determined
by the occupied s.p. levels (\textit{i.e.} HF configuration),
as exemplified in Ref.~\cite{ref:SNM16} and further confirmed in this study.
To draw $E(q_0)$, \textit{i.e.} energy curve as a function of $q_0$,
the constrained HF (CHF) calculations have been carried out.
The procedures for the CHF calculations are described in Ref.~\cite{ref:SNM16}.
Contribution of the tensor force is evaluated
by $E^{(\mathrm{TN})}(q_0)=\langle\Phi(q_0)|\hat{v}^{(\mathrm{TN})}|\Phi(q_0)\rangle$,
where $\hat{v}^{(\mathrm{TN})}$ is the tensor force
and $|\Phi(q_0)\rangle$ represents the CHF state.
We shall compare energies $E(q_0)$, $E(q_0)-E^{(\mathrm{TN})}(q_0)$,
both of which are obtained from the M3Y-P6 interaction,
and $E(q_0)$ obtained from the Gogny-D1M interaction.
It is again emphasized that $\hat{v}^{(\mathrm{TN})}$ in M3Y-P6 is realistic,
which has been derived via the $G$-matrix
without adjusting to experimental data,
and that this tensor force reproduces
variation of relative s.p. energies of $p0d_{3/2}$ and $p1s_{1/2}$
from $^{40}$Ca to $^{48}$Ca remarkably well~\cite{ref:NSM13}.
The s.p. energy $\varepsilon(\nu)$ at the minima will be shown as well,
where $\nu$ denotes the s.p. level in the HF.
The s.p. energies are useful for analyzing configurations that yield the minima.
To visualize contribution of the tensor force to $\varepsilon(\nu)$,
$\varepsilon(\nu)-\varepsilon^{(\mathrm{TN})}(\nu)$ is also displayed,
where $\varepsilon^{(\mathrm{TN})}(\nu)
=2\sum_{\nu'\,(>0)} n_{\nu'}\langle\nu\nu'|v^{(\mathrm{TN})}|\nu\nu'\rangle$
with the occupation probability $n_\nu$.

We here recall several properties of the tensor force at the MF level,
which have been established in Refs.~\cite{ref:Vtn,ref:SNM16,ref:Sky-TNS}.
\begin{itemize}
 \item[i)] The tensor force primarily provides proton-neutron correlations.
 \item[ii)] The tensor force acts repulsively.
 \item[iii)] Tensor-force effects are perturbative at the HF level,
   but configuration-dependent.
   For a fixed configuration, $E^{(\mathrm{TN})}(q_0)$ is insensitive to $q_0$.
 \item[iv)] The tensor force tends to lower the spherical state
   relative to the deformed ones at the $\ell s$-closed magic numbers,
   while the opposite holds at the $jj$-closed magic numbers.
\end{itemize}
The point iii) allows us to analyze tensor-force effects
in terms of the spherical orbits.
The tensor force acts repulsively (attractively)
on neutron $j=\ell+1/2$ ($j=\ell-1/2$) orbits
as a proton $j=\ell+1/2$ orbit is occupied~\cite{ref:Vtn}.
This accounts for the point ii),
because a valence $j=\ell+1/2$ orbit should have higher occupation probability
than its $\ell s$ partner,
both for protons and neutrons~\cite{ref:SNM16}.
The point iv) takes place
because the tensor force works when the spin d.o.f. are active.
Its contribution depends on how well the spin d.o.f.
are saturated~\cite{ref:SNM16,ref:Sky-TNS}.
As $Z=40$ forms an $\ell s$-closed shell at the spherical limit,
it is expected that the tensor force tends to favor sphericity in the Zr nuclei.
Owing to the point iii), the HF framework is advantageous
in investigating tensor-force effects~\cite{ref:SNM16},
and therefore indispensable in reconstructing the MF scheme
including the tensor force.
Mixing of the HF configurations,
which occurs via correlations beyond HF (\textit{e.g.} the pairing),
leads to obscurity in viewing the tensor-force effects.

\subsection{$^{90}$Zr}\label{subsec:N50}

\begin{figure}
\includegraphics[scale=0.7]{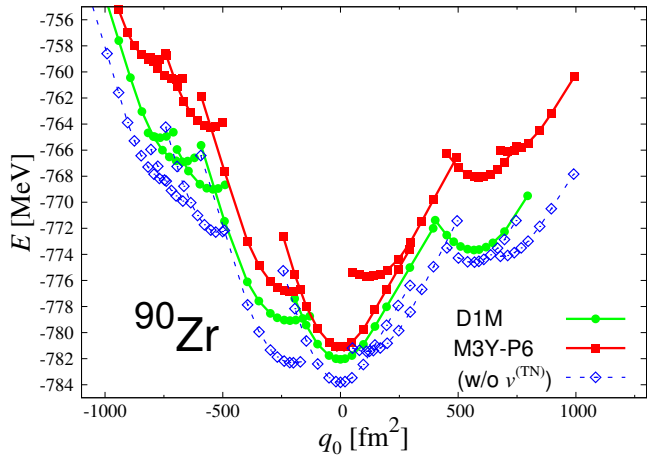}
\caption{(Color online) CHF results
of $E(q_0)$ (red squares)
and $E(q_0)-E^{(\mathrm{TN})}(q_0)$ (blue open diamonds)
for $^{90}$Zr, which are obtained with M3Y-P6.
$E(q_0)$ with D1M (green circles) are also plotted.
Lines are drawn to guide eyes.
For reference, energy obtained by the spherical HFB calculation~\cite{ref:NS14}
is shown by the red cross,
though invisible here since it is merged with the HF energy at $q_0=0$.
\label{fig:Zr90_E-q0}}
\end{figure}

The energy curve $E(q_0)$ for $^{90}$Zr is shown in Fig.~\ref{fig:Zr90_E-q0}.
While several minima are observed on the prolate and the oblate sides,
the HF energy becomes lowest at $q_0=0$.
It is confirmed from the spherical HFB energy,
which is not distinguishable from the HF energy at $q_0=0$
in Fig.~\ref{fig:Zr90_E-q0},
that the pair correlation is so weak at $q_0=0$.
This spherical minimum is well developed both in the M3Y-P6 and the D1M results,
compared to $E-E^{(\mathrm{TN})}$.
Comparing the full M3Y-P6 energy $E$ with $E-E^{(\mathrm{TN})}$,
we confirm the points raised above.
The tensor force acts repulsively,
and $E^{(\mathrm{TN})}$ does not change much
along the curve connected to each local minimum.
Moreover, the tensor force helps the spherical configuration
to be the distinct absolute minimum,
as $E^{(\mathrm{TN})}$ comes larger as the deformation develops.
On account of this tensor-force effect,
it can be considered that, in the D1M interaction,
the tensor-force effects are partly incorporated
into the other channels in an effective manner,
so that the property of $^{90}$Zr should be reasonably reproduced.

\begin{figure}
\hspace*{-2cm}\includegraphics[scale=0.7]{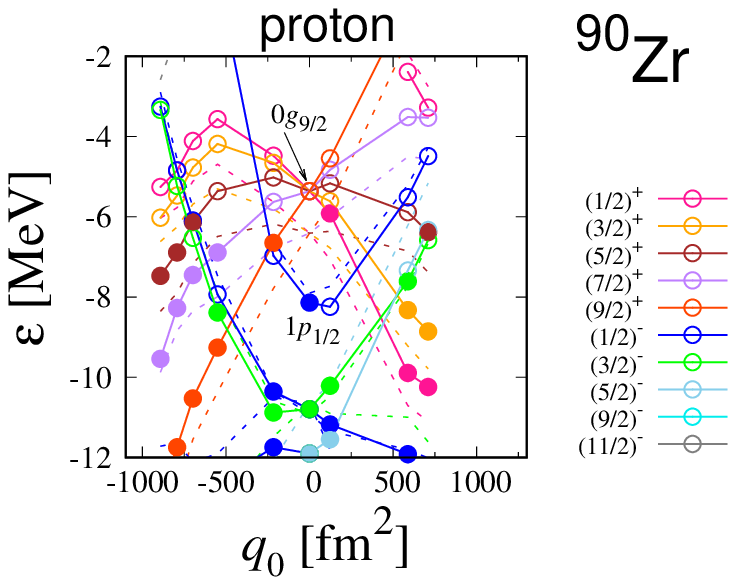}
\hspace*{-0.3cm}\includegraphics[scale=0.7]{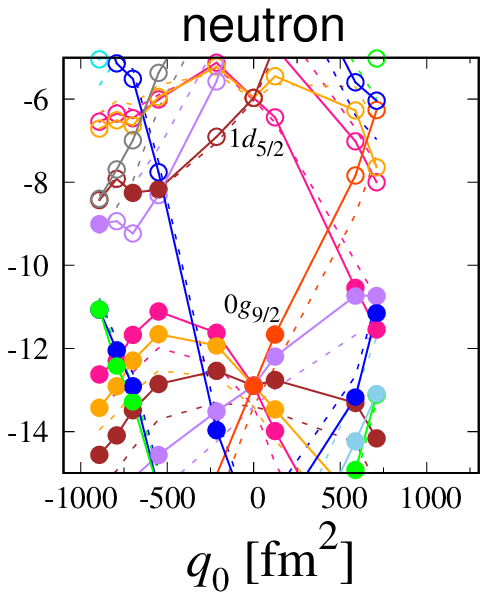}\hspace*{-4cm}
\vspace*{-1cm}
\caption{(Color online) Proton and neutron s.p. energies $\varepsilon(\nu)$
in $^{90}$Zr obtained by the axial HF calculations with M3Y-P6,
at the minima shown in Fig.~\ref{fig:Zr90_E-q0}.
Occupied (unoccupied) levels are represented by the filled (open) circles
connected by the solid lines.
Quantum number of each level $\Omega^\pi$ is distinguished by colors,
and is indicated in the middle.
Dashed lines show $\varepsilon(\nu)-\varepsilon^{(\mathrm{TN})}(\nu)$.
Labels for several spherical orbits are given for reference.
\label{fig:Zr90_spe}}
\end{figure}

Figure~\ref{fig:Zr90_spe} depicts the s.p. levels around the Fermi energy
at individual local minima,
which are obtained by the HF calculation with M3Y-P6.
The occupied levels are presented by the filled circles.
To examine the tensor-force effects,
$[\varepsilon(\nu)-\varepsilon^{(\mathrm{TN})}(\nu)]$'s
are shown by the dashed lines.
Since the $N=50$ shell gap is large,
neutrons are hardly excited at relatively small $|q_0|$.
What triggers deformation is
excitation of protons from the $1p_{1/2}$ orbit to $0g_{9/2}$.
The tensor force enlarges the shell gap between these orbits,
which lowers the energy of the spherical minimum relative to the deformed ones.
In addition, it is seen in Fig.~\ref{fig:Zr90_spe}
that $\varepsilon$ changes more rapidly
than $\varepsilon-\varepsilon^{(\mathrm{TN})}$,
as the configuration (\textit{i.e.} $q_0$ in the figure) varies.
This effect is linked to the degree of the spin saturation.
Owing to the $\ell s$ closure for protons,
contribution of the tensor force comes small at the spherical limit.
As the deformation develops,
the spin saturation is lost in the proton state, 
giving rise to the larger $E^{(\mathrm{TN})}$ in Fig.~\ref{fig:Zr90_E-q0}.

\subsection{$^{80}$Zr}\label{subsec:N40}

\begin{figure}
\includegraphics[scale=0.7]{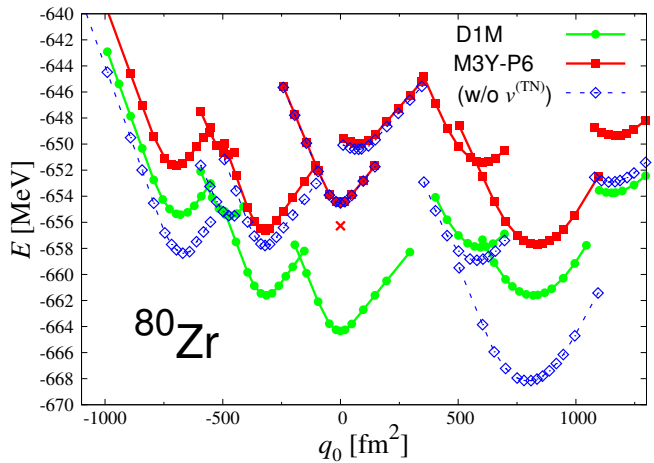}
\caption{(Color online) CHF results for $^{80}$Zr.
See Fig.~\protect\ref{fig:Zr90_E-q0} for conventions.
\label{fig:Zr80_E-q0}}
\end{figure}

We next turn to $^{80}$Zr.
See Fig.~\ref{fig:Zr80_E-q0} for the energy curve.
Because of $Z=N=40$, $^{80}$Zr would be spherical
if the shell gap between $pf$-shell orbits and $0g_{9/2}$ were large.
In practice, the lowest energy is obtained at $q_0=0$
by the HF calculation with D1M.
The same holds in the HFB result with D1S~\cite{ref:D1S-Web}.
In contrast, in the M3Y-P6 result
a prolate and an oblate minima lie lower than the spherical minimum.
Whereas the pairing lowers the energy of the spherical state,
the deformed minima are even lower than the spherical HFB energy
(the red cross in Fig.~\ref{fig:Zr80_E-q0}).
The lowest energy is given by the prolate minimum
with $q_0\approx 800\,\mathrm{fm}^2$,
which corresponds to $\beta\approx 0.5$.
The difference in the deformation between M3Y-P6 and D1M
is traced back to the $N$-dependence of the shell gap
as discussed in Ref.~\cite{ref:NS14}.
Even though the tensor-force effects could be partly incorporated in D1M
in an effective manner,
it could be difficult to mimic all aspects of them,
particularly in the case that the tensor force directly affects
the $Z$- or $N$-dependence of the shell gap.
The D1M interaction reproduces properties of $^{90}$Zr
without the tensor force,
and this seems to make it difficult to describe the deformation of $^{80}$Zr
at the MF level.

In $E-E^{(\mathrm{TN})}$,
the prolate minimum has much lower energy than the other minima.
Because of the $\ell s$ closure both for protons and neutrons,
$E^{(\mathrm{TN})}$ is negligibly small at the spherical minimum.
The repulsive effect of the tensor force is the stronger
for the larger deformation,
and diminishes energy difference between the spherical and the prolate minima.
Still, the prolate minimum stays lower than the spherical minimum
in the full M3Y-P6 result.
Though to less degree, the same mechanism works for the oblate minimum,
also staying lower than the spherical minimum.
The prolate and the oblate states are energetically competing
in the full M3Y-P6 result.
The close energies of the prolate and the oblate states
suggest shape coexistence at low energy.

It is mentioned that, as illustrated by the D1M results,
the deformation at $^{80}$Zr has not been easy
to be reproduced with the interactions that do not contain the tensor force,
until beyond-MF effects are taken account of~\cite{ref:RE11}.
Though there exist exceptions
(\textit{e.g.} the HF result with SkM$^\ast$ in Table~\ref{tab:pred-shape}),
$^{80}$Zr is hardly deformed also with the Skyrme interactions at the MF level.
If we use an interaction
by which the $Z=40$ shell gap is kept large at $^{90}$Zr,
this gap becomes even larger at $N=40$ if we do not have the tensor force.
The realistic tensor force reverses this trend,
as illustrated in Fig.~9 of Ref.~\cite{ref:NS14}.
On the other hand,
there is another effect of the tensor force giving the opposite tendency;
namely, favoring the spherical shape.
The present results imply that the semi-realistic interaction
yields the tensor-force effects with an appropriate balance.

\begin{figure}
\hspace*{-2cm}\includegraphics[scale=0.7]{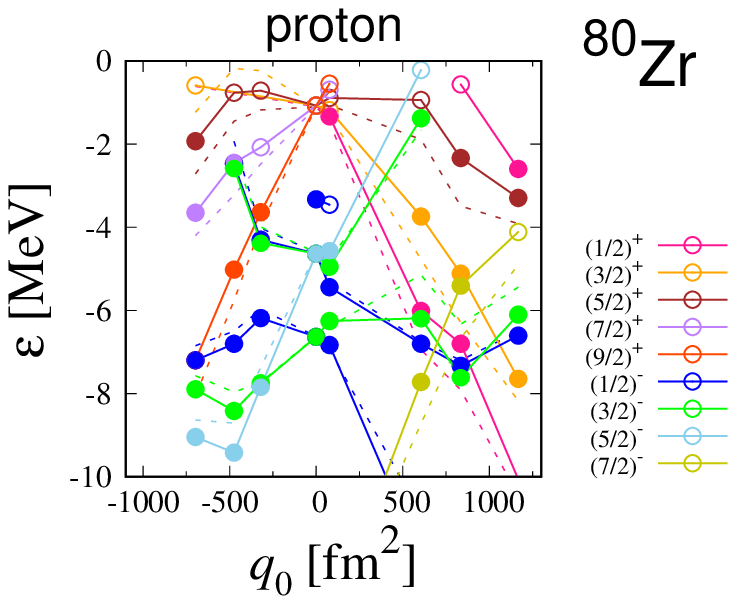}
\hspace*{-0.3cm}\includegraphics[scale=0.7]{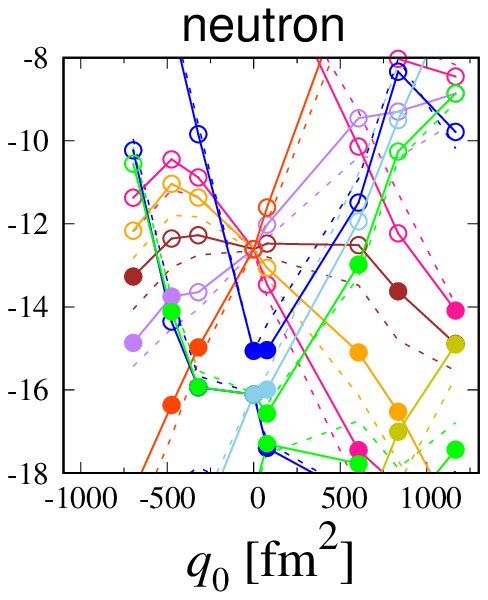}\hspace*{-4cm}
\vspace*{-1cm}
\caption{(Color online) Proton and neutron s.p. energies in $^{80}$Zr
obtained from the HF calculations with M3Y-P6,
at the minima shown in Fig.~\ref{fig:Zr80_E-q0}.
See Fig.~\ref{fig:Zr90_spe} for conventions.
\label{fig:Zr80_spe}}
\end{figure}

The s.p. levels are shown in Fig.~\ref{fig:Zr80_spe}.
In this nucleus, deformation is driven
by excitation of either protons or neutrons
from the $pf$-shell orbitals to $0g_{9/2}$.
At the lowest minimum with $q_0\approx 800\,\mathrm{fm}^2$,
four protons and four neutrons are excited
to the levels coming down from $0g_{9/2}$,
indicating that it is basically an $8p$-$8h$ state
in terms of the spherical orbitals.
The oblate minimum with $q_0\approx -300\,\mathrm{fm}^2$,
which is the second lowest in energy,
has a $4p$-$4h$ configuration.

\subsection{$^{96}$Zr}\label{subsec:N56}

\begin{figure}
\includegraphics[scale=0.7]{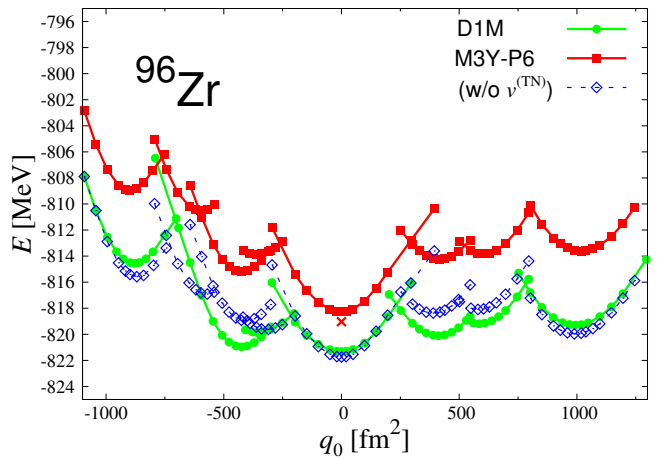}
\caption{(Color online) CHF results for $^{96}$Zr.
See Fig.~\protect\ref{fig:Zr90_E-q0} for conventions.
\label{fig:Zr96_E-q0}}
\end{figure}

In Ref.~\cite{ref:NS14},
$N=56$ has been predicted to be submagic at $^{96}$Zr,
based on the quenched pair correlation in the spherical HFB result.
This submagic nature is in accordance
with the high $E_x(2^+_1)$ in measurement~\cite{ref:NuDat},
and has been accounted for by the enhanced shell gap owing to the tensor force.
Figure~\ref{fig:Zr96_E-q0} supports this prediction.
Even after the deformation is taken into account,
M3Y-P6 gives the lowest energy at $q_0=0$,
whose energy is distinctly lower than the deformed local minima.
As seen in Fig.~\ref{fig:Zr96_E-q0},
the energy difference between HF and HFB at $q_0=0$ is tiny though visible.
Although the spherical configuration is the lowest also in the D1M result,
energy difference between the spherical and deformed minima is small
and correlations beyond HF could invert them.
In the HFB result with D1S~\cite{ref:D1S-Web},
a shallow minimum around $q_0=0$ was reported.

\begin{figure}
\hspace*{-2cm}\includegraphics[scale=0.7]{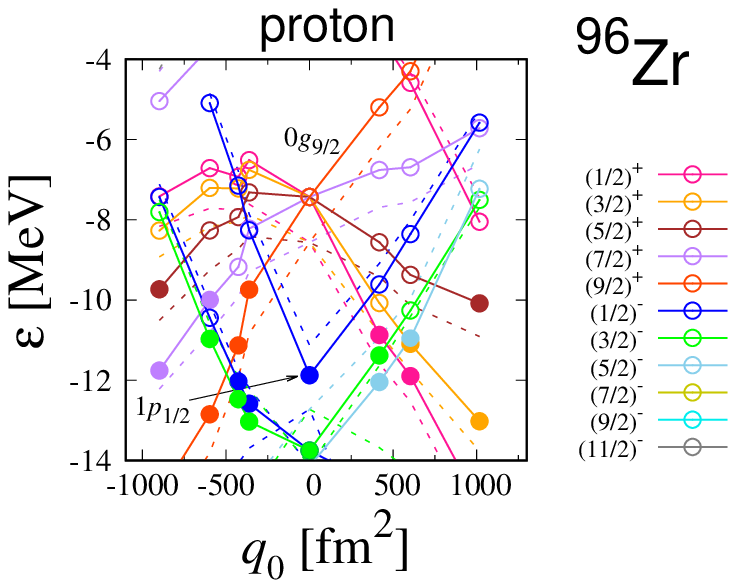}
\hspace*{-0.3cm}\includegraphics[scale=0.7]{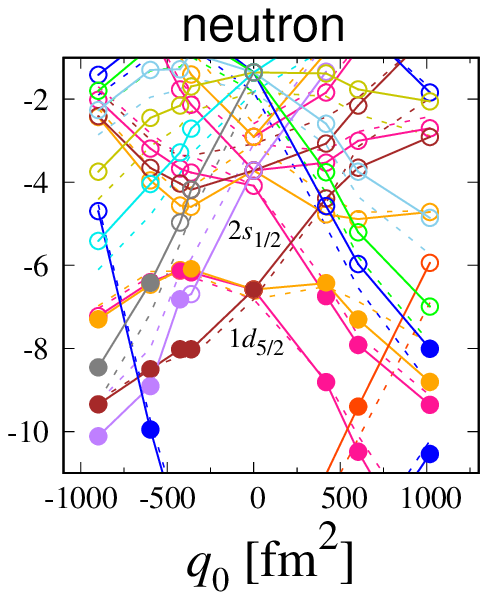}\hspace*{-4cm}
\vspace*{-1cm}
\caption{(Color online) Proton and neutron s.p. energies in $^{96}$Zr
obtained from the HF calculations with M3Y-P6,
at the minima shown in Fig.~\ref{fig:Zr96_E-q0}.
See Fig.~\ref{fig:Zr90_spe} for conventions.
\label{fig:Zr96_spe}}
\end{figure}

The s.p. levels in Fig.~\ref{fig:Zr96_spe} further establish
the role of the tensor force in the magicity of $^{96}$Zr
discussed in Ref.~\cite{ref:NS14}.
For $Z=40$, the shell gap between $p1p_{1/2}$ and $p0g_{9/2}$ is enhanced
by the tensor force combined with occupation of $n0g_{9/2}$ and $n1d_{5/2}$.
For $N=56$, the gap between $n1d_{5/2}$ and $n0g_{7/2}$ remains relatively large.
Thereby $n2s_{1/2}$ becomes the lowest unoccupied level.
Similar crossing of the spherical orbitals
has been pointed out for the Ni isotopes~\cite{ref:Nak10b}.
As presented in Figs. 6 and 9 of Ref.~\cite{ref:NS14},
D1M gives smaller shell gaps than M3Y-P6 both for protons and neutrons,
by which the energies of the deformed minima
come close to that of the spherical minimum.

We here discuss effects of the tensor force
under the presence of a unique-parity orbit.
Because the $\ell s$ partner of the unique-parity orbit has much higher energy,
the system becomes away from the spin saturation
as the unique-parity orbit is occupied.
This makes tensor-force effects stronger,
as is observed for the $q_0\approx 1000\,\mathrm{fm}^2$ minimum
in Fig.~\ref{fig:Zr96_E-q0}.
In the lower part of the major shell (\textit{e.g.} $N=50-82$),
the unique-parity orbit (\textit{e.g.} $n0h_{11/2}$) is not occupied
at the spherical limit,
while becomes occupied to a greater extent as the deformation grows.
Then the tensor force, which is repulsive, acts more strongly.
Therefore, if the unique-parity orbit is located
above but not distant from the Fermi energy,
the tensor force tends to lower energy of the spherical configurations
relative to those of the deformed ones.
As a result,
the $q_0\approx 1000\,\mathrm{fm}^2$ state lies significantly higher
than the spherical minimum
and the doubly-magic nature is enhanced in $^{96}$Zr,
as recognized by comparing $E$ and $E-E^{(\mathrm{TN})}$
in Fig.~\ref{fig:Zr96_E-q0}.
In the upper part of the major shell (\textit{i.e.} for larger $N$),
the unique-parity orbit is already occupied at the spherical limit
to a certain extent.
Therefore this mechanism is expected to work weaker
than in the lower part of the major shell.

\subsection{$^{100}$Zr}\label{subsec:N60}

Experiments have indicated well-deformed ground states in $N\geq 60$.
It has been recently argued
that the sudden shape change from $^{98}$Zr to $^{100}$Zr
may be interpreted as a quantum phase transition.
It deserves having a close look at the shape evolution
in this particular region,
and we take $^{100}$Zr as an example.

\begin{figure}
\includegraphics[scale=0.7]{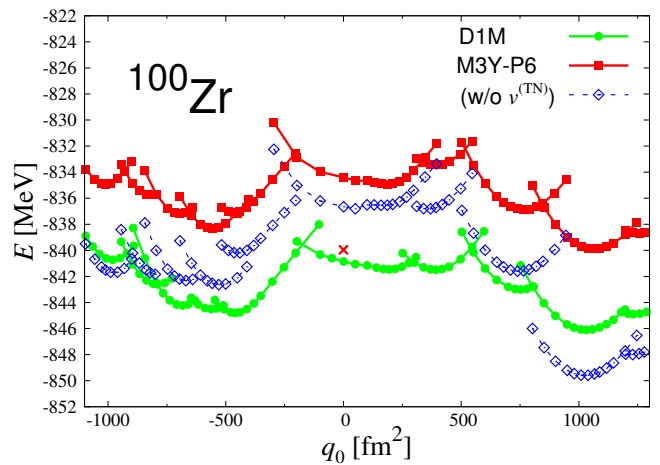}
\caption{(Color online) CHF results for $^{100}$Zr.
See Fig.~\protect\ref{fig:Zr90_E-q0} for conventions.
\label{fig:Zr100_E-q0}}
\end{figure}

\begin{figure}
\hspace*{-2cm}\includegraphics[scale=0.7]{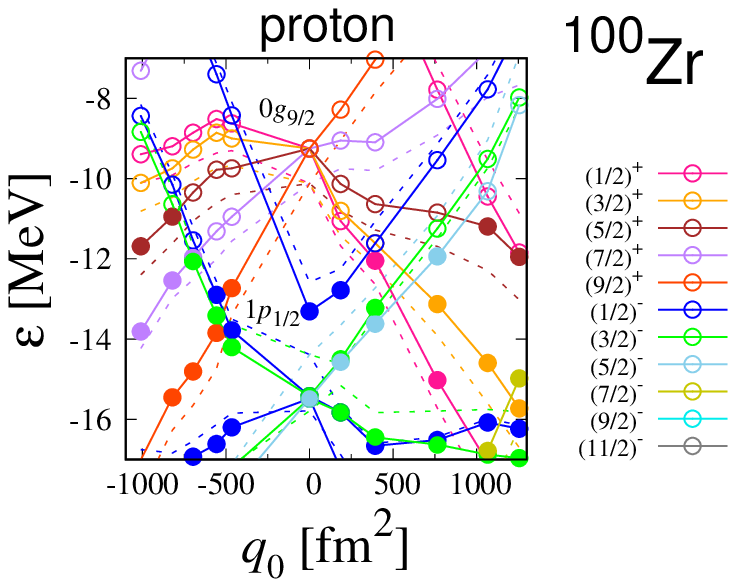}
\hspace*{-0.3cm}\includegraphics[scale=0.7]{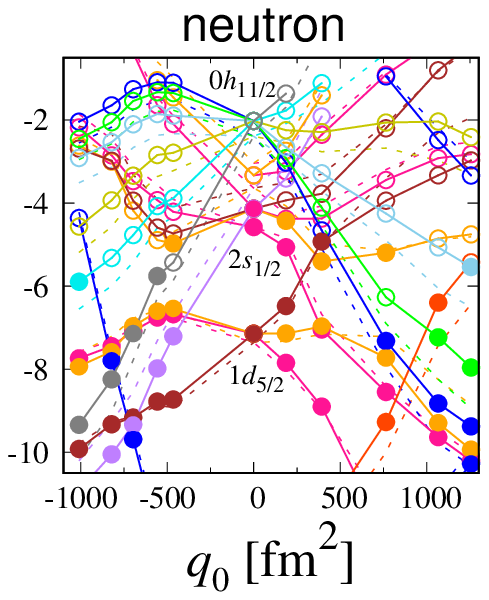}\hspace*{-4cm}
\vspace*{-1cm}
\caption{(Color online) Proton and neutron s.p. energies in $^{100}$Zr
obtained from the HF calculations with M3Y-P6,
at the minima shown in Fig.~\ref{fig:Zr100_E-q0}.
The levels at $q_0=0$ are the results of the spherical HF calculation.
See Fig.~\ref{fig:Zr90_spe} for conventions.
\label{fig:Zr100_spe}}
\end{figure}

Figure~\ref{fig:Zr100_E-q0} shows the energy curve for $^{100}$Zr.
Both in the M3Y-P6 and D1M results,
the lowest energy is obtained by the well-deformed prolate state
with $q_0\approx 1100\,\mathrm{fm}^2$.
Whereas this lowest energy is close
to the spherical HFB energy~\cite{ref:NS14}
with the difference less than $0.1\,\mathrm{MeV}$,
it is likely that this prolate state is distinctly lower
after the pairing and the rotational correlations are taken into account.
Between the lowest state and the local minimum closest to $q_0=0$,
we observe two other local minima.
Several minima are found on the oblate side as well.
Figure~\ref{fig:Zr100_spe} tells us detailed structure at these minima.
While we do not have a minimum at $q_0=0$ in the axial HF calculations,
the s.p. levels obtained from the spherical HF calculation
are presented for reference.

We find energy minima at similar $q_0$'s between $^{96}$Zr and $^{100}$Zr,
although their energies are shifted depending on the configurations.
It is interesting to compare the configurations at the minima
corresponding to each other, between these nuclei.
At the minima for increasing $|q_0|$,
protons in the $pf$-shell are excited to $0g_{9/2}$ two by two,
and the minima are characterized by the proton configurations in this region.
In practice, the first s.p. level dominated by $p0g_{9/2}$,
which has $\Omega^\pi=(1/2)^+$,
is occupied at the minimum slightly below $q_0=500\,\mathrm{fm}^2$,
the second having $\Omega^\pi=(3/2)^+$ above $q_0=500\,\mathrm{fm}^2$,
and the third having $\Omega^\pi=(5/2)^+$ at $q_0\approx 1000\,\mathrm{fm}^2$
for both nuclei.
Analogous variation of proton configurations is seen at the oblate minima.
Then neutron configurations at the minima and their relative energies
determine which minimum provides the lowest energy.
Whereas the even-parity levels near the Fermi energy mix one another,
the unique-parity orbit $n0h_{11/2}$ stays nearly pure,
and occupation number on $n0h_{11/2}$ carries important structural information.
Owing to the four more neutrons,
$n0h_{11/2}$ is closer to the highest occupied level
in $^{100}$Zr than in $^{96}$Zr at the spherical limit.
This facilitates crossing of the s.p. levels
one of which is dominated by $n0h_{11/2}$,
particularly on the prolate side.
Two neutrons occupy $0h_{11/2}$ at $q_0\approx 800\,\mathrm{fm}^2$ in $^{100}$Zr,
while at $q_0\approx 1000\,\mathrm{fm}^2$ in $^{96}$Zr.
At the $q_0\approx 1100\,\mathrm{fm}^2$ minimum of $^{100}$Zr,
four neutrons occupy the levels corresponding to $0h_{11/2}$.

The repulsive effect of the tensor force is strong
at the $q_0\approx 1100\,\mathrm{fm}^2$ state
because of the occupation of $n0h_{11/2}$,
whereas this state remains to yield the lowest energy,
diminishing the energy difference between this and the other minima.
This effect becomes more important in the structure of $^{98}$Zr.
Although the lowest minimum has prolate shape also in $^{98}$Zr
in the present HF calculation with M3Y-P6,
it is almost degenerate with an oblate minimum.
The energy difference between the prolate and the oblate minima
is thinner ($0.5\,\mathrm{MeV}$) in $^{98}$Zr,
which is compared to $1.5\,\mathrm{MeV}$ in $^{100}$Zr
shown in Fig.~\ref{fig:Zr100_E-q0}.
Because of this small difference,
these two minima might be inverted by correlations beyond HF,
while such inversion is unlikely to take place in $E-E^{(\mathrm{TN})}$.
We also mention that the spherical HFB energy at $^{98}$Zr~\cite{ref:NS14}
is lower than the lowest axial HF energy by $1.8\,\mathrm{MeV}$.

\subsection{$^{104}$Zr}\label{subsec:N64}

\begin{figure}
\includegraphics[scale=0.7]{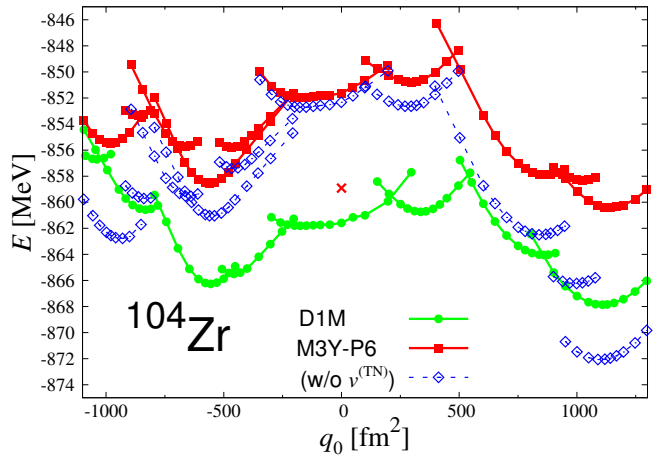}
\caption{(Color online) CHF results for $^{104}$Zr.
See Fig.~\protect\ref{fig:Zr90_E-q0} for conventions.
\label{fig:Zr104_E-q0}}
\end{figure}

In $60\leq N\leq 70$, measured $E_x(2^+_1)$'s
are steadily low~\cite{ref:Che70,ref:NuDat,ref:Sum11,ref:Pau17,ref:Hot91}.
Measured spectroscopic properties are well described
by the prolate deformation up to $^{104}$Zr~\cite{ref:Liu11}.
It has been claimed that $^{104}$Zr has the highest collectivity
among the Zr isotopes,
because measured $E_x(2^+_1)$ is the lowest
and $B(E2)$ is the strongest~\cite{ref:Hwa06,ref:Bro15}.
We next pick up this nucleus.
As seen in Fig.~\ref{fig:q0-N},
the $q_0$ value at the g.s. does not change much in $58\leq N\leq 72$
in the present calculations with M3Y-P6.
It is observed in Fig.~\ref{fig:Zr104_E-q0}
that the lowest minimum is well developed in $^{104}$Zr,
having $q_0\approx 1100\,\mathrm{fm}^2$,
and the second minimum is found at $q_0\approx -600\,\mathrm{fm}^2$.
It is noticed that these $q_0$ values are close to those in $^{100}$Zr
shown in Fig.~\ref{fig:Zr100_E-q0}.
Typified by these two minima, $E(q_0)$ in $^{104}$Zr has a similarity
to $E(q_0)$ in $^{100}$Zr,
as the $q_0$ values giving the local minima are not very different.

\begin{figure}
\hspace*{-2cm}\includegraphics[scale=0.7]{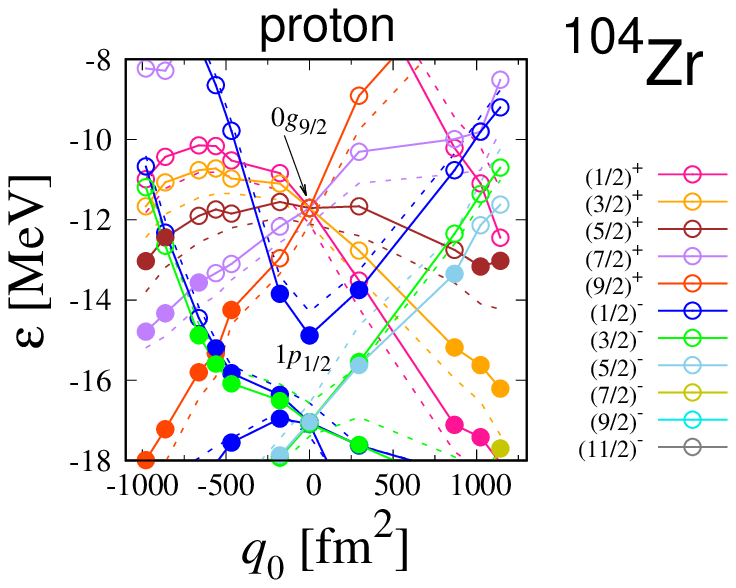}
\hspace*{-0.3cm}\includegraphics[scale=0.7]{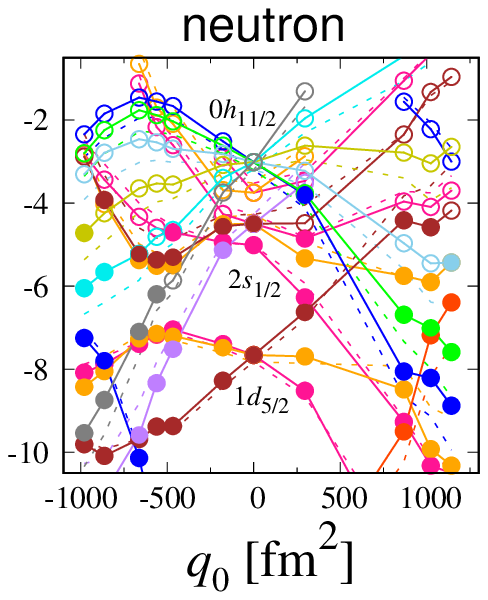}\hspace*{-4cm}
\vspace*{-1cm}
\caption{(Color online) Proton and neutron s.p. energies in $^{104}$Zr
obtained from the HF calculations with M3Y-P6,
at the minima shown in Fig.~\ref{fig:Zr104_E-q0}.
See Fig.~\ref{fig:Zr90_spe} for conventions.
\label{fig:Zr104_spe}}
\end{figure}

The s.p. levels at the minima of $^{104}$Zr are depicted
in Fig.~\ref{fig:Zr104_spe}.
At the absolute minimum with $q_0\approx 1100\,\mathrm{fm}^2$,
six neutrons occupy the levels connected to $0h_{11/2}$.
Half of the levels belonging to $n0h_{11/2}$ (the $\Omega\leq 5/2$ levels)
come down as the prolate deformation grows,
while the other half go up,
and the lower half of these levels are occupied at the minimum.
At the second minimum which lies on the oblate side,
two neutrons occupy the level connected to $0h_{11/2}$.
The difference in the occupation on $n0h_{11/2}$
is reflected by $E^{(\mathrm{TN})}$ in Fig.~\ref{fig:Zr104_E-q0}.

\subsection{$^{114}$Zr}\label{subsec:N74}

\begin{figure}
\includegraphics[scale=0.7]{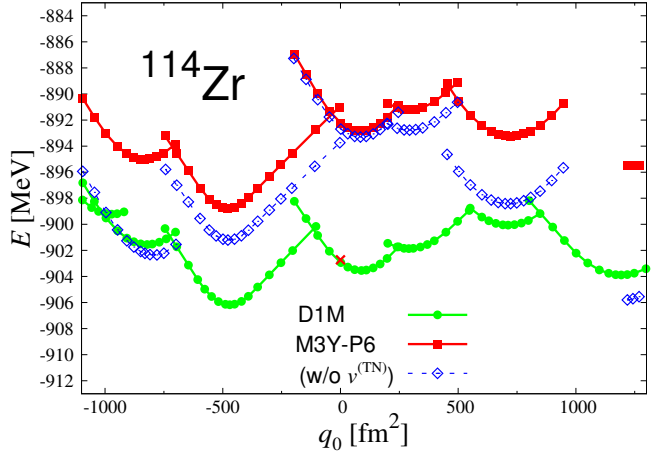}
\caption{(Color online) CHF results for $^{114}$Zr.
See Fig.~\protect\ref{fig:Zr90_E-q0} for conventions.
\label{fig:Zr114_E-q0}}
\end{figure}

In Fig.~\ref{fig:q0-N},
we have found a sudden shape transition from prolate to oblate at $^{114}$Zr.
The oblate state has $q_0\approx -500\,\mathrm{fm}^2$.
The energy curve for $^{114}$Zr is exhibited in Fig.~\ref{fig:Zr114_E-q0}.
The tensor force plays a crucial role in the shape change.
If there were no tensor force in the effective interaction,
the prolate state with $q_0\approx 1300\,\mathrm{fm}^2$
should be lower than the oblate minimum
as recognized from $E-E^{(\mathrm{TN})}$.
However, the repulsion due to the tensor force is so strong
at the prolate minimum
that the energy of this state could become higher than the oblate state.

\begin{figure}
\hspace*{-2cm}\includegraphics[scale=0.7]{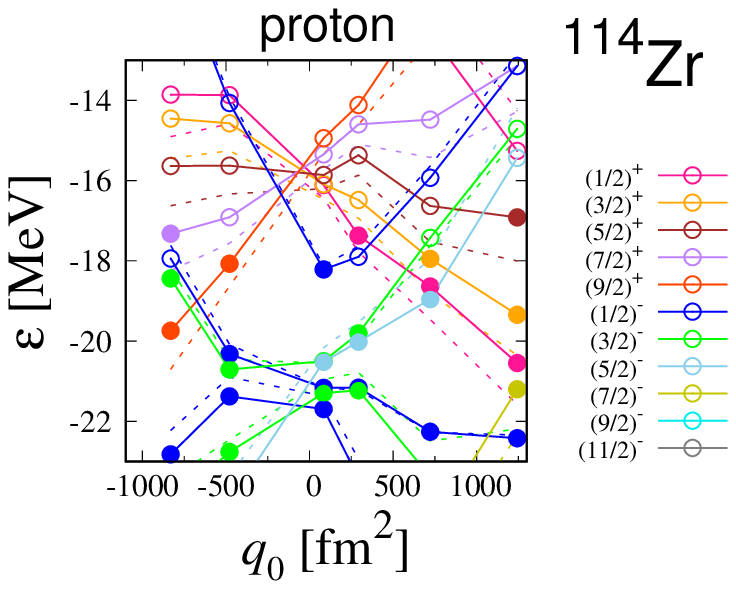}
\hspace*{-0.3cm}\includegraphics[scale=0.7]{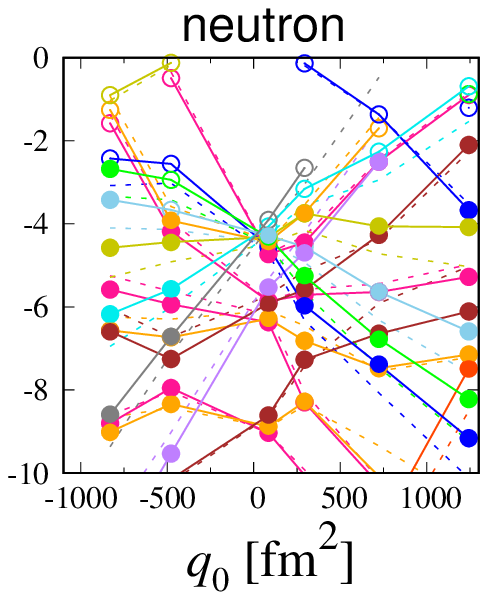}\hspace*{-4cm}
\vspace*{-1cm}
\caption{(Color online) Proton and neutron s.p. energies in $^{114}$Zr
obtained from the HF calculations with M3Y-P6,
at the minima shown in Fig.~\ref{fig:Zr114_E-q0}.
See Fig.~\ref{fig:Zr90_spe} for conventions.
\label{fig:Zr114_spe}}
\end{figure}

The occupied s.p. levels shown in Fig.~\ref{fig:Zr114_spe} are useful
in anatomizing the shape change in this nucleus.
It should be noticed that four neutrons occupy $0h_{11/2}$
at the spherical limit, because of $N=74$.
At the absolute minimum located on the oblate side,
six neutrons occupy the levels connected to $0h_{11/2}$.
The repulsion of the tensor force is stronger
than at the spherical configuration, but not so significantly.
On the contrary, at the prolate minimum with $q_0\approx 1300\,\mathrm{fm}^2$,
eight neutrons occupy the levels connected to $0h_{11/2}$.
Moreover, two more neutrons occupy an odd-parity level
coming down from the $n1f_{7/2}$ orbit in the upper major shell,
which further enhance the repulsive tensor-force effect.

The spherical HFB calculation yields lower energy than the deformed HF energies.
As presented in Table~\ref{tab:pred-shape},
the HFB calculations with D1S and SLy4 predict spherical shape at $^{114}$Zr.
It should be postponed to address full MF prediction with M3Y-P6
on the shape of $^{114}$Zr,
until the deformation and the pairing are simultaneously taken into account.

\subsection{$^{120}$Zr}\label{subsec:N80}

In the present calculation
the shape returns to be almost spherical at $^{120}$Zr,
as shown in Fig.~\ref{fig:q0-N}.
It is found that a state in the vicinity of $q_0=0$ is the lowest
both in the M3Y-P6 and the D1M results.
However, if $E^{(\mathrm{TN})}$ is subtracted,
the oblate state with $q_0\approx -500\,\mathrm{fm}^2$ stays lowest.
The good doubly magic nature has been predicted for $^{122}$Zr
in Ref.~\cite{ref:NS14},
which is preserved after the deformation is taken into account.
In $E-E^{(\mathrm{TN})}$ in $^{122}$Zr,
the oblate minimum lies relatively close to the spherical configuration,
with the difference less than $3\,\mathrm{MeV}$.
The doubly-magic nature of $^{122}$Zr is enhanced
by the tensor force to a significant extent.

\begin{figure}
\includegraphics[scale=0.7]{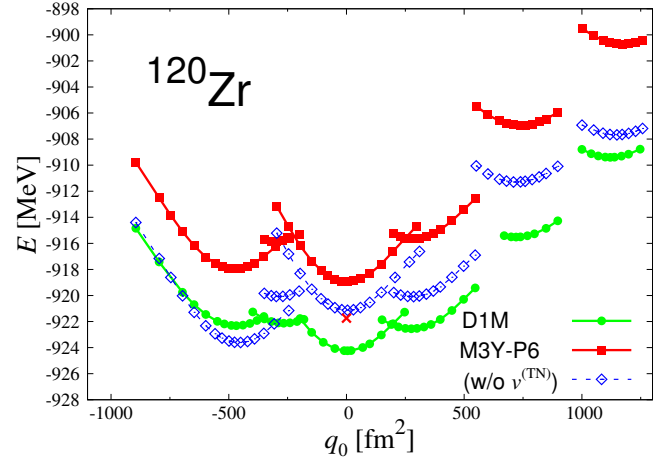}
\caption{(Color online) CHF results for $^{120}$Zr.
See Fig.~\protect\ref{fig:Zr90_E-q0} for conventions.
\label{fig:Zr120_E-q0}}
\end{figure}

\begin{figure}
\hspace*{-2cm}\includegraphics[scale=0.7]{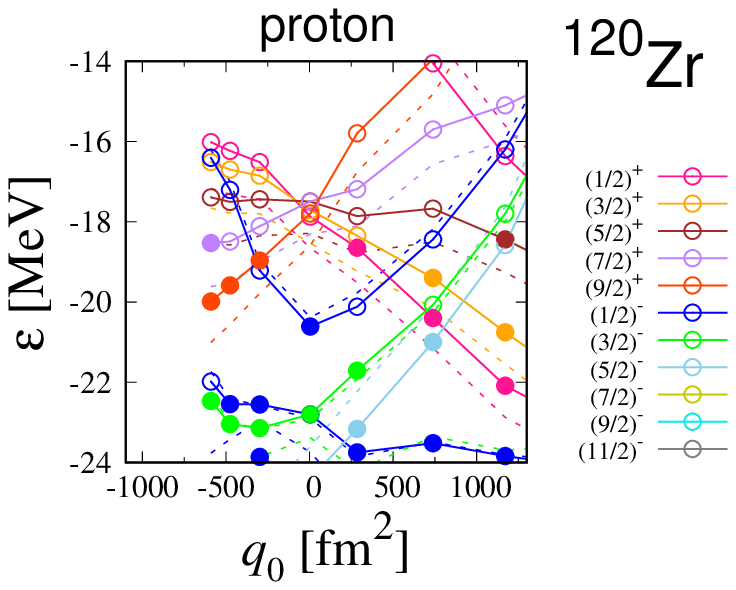}
\hspace*{-0.3cm}\includegraphics[scale=0.7]{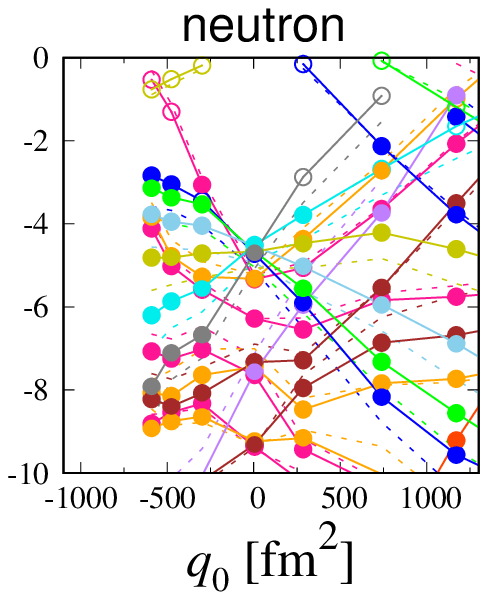}\hspace*{-4cm}
\vspace*{-1cm}
\caption{(Color online) Proton and neutron s.p. energies in $^{120}$Zr
obtained from the HF calculations with M3Y-P6,
at the minima shown in Fig.~\ref{fig:Zr120_E-q0}.
See Fig.~\ref{fig:Zr90_spe} for conventions.
\label{fig:Zr120_spe}}
\end{figure}

The effect of the tensor force is traced back
to the s.p. energies presented in Fig.~\ref{fig:Zr120_spe},
as typically observed in $\varepsilon$
and $\varepsilon-\varepsilon^{(\mathrm{TN})}$
of the proton $\Omega^\pi=(9/2)^+$ level.
Because of the energy gain of this level,
the oblate configuration becomes lowest in $E-E^{(\mathrm{TN})}$.
However, the tensor force raised this level significantly,
finally making the spherical configuration the lowest
in Fig.~\ref{fig:Zr120_E-q0}.
Thus, occupation of $p0g_{9/2}$ is primarily responsible for the difference
in $E^{(\mathrm{TN})}$.
Role of the $n0h_{11/2}$ orbit is not quite conspicuous for this nucleus,
since this orbit is mostly occupied even at the spherical limit.
It is found that excitation across the $N=82$ spherical shell gap
hardly takes place on the oblate side,
while excitation to $n1f_{7/2}$ may occur on the prolate side.

\section{Discussion and summary\label{sec:discuss}}

In Ref.~\cite{ref:NS14}, magic numbers are well predicted
by the spherical MF calculations with semi-realistic interaction M3Y-P6.
Whereas no deformation d.o.f. were handled,
the pairing was used as a representative of correlations.
The results were in accordance with the available data for most nuclei
all over the nuclear chart.
This suggests that the description of the nuclear structure
may be rather simplified
by the semi-realistic interaction containing the realistic tensor force,
as a good first approximation is obtained within the MF framework.
However, a discrepancy had been found in the $Z=40$ magicity
in the neutron-rich Zr nuclei.
Although it has been known that the Zr nuclei are deformed
in $60\leq N\lesssim 70$,
the spherical HFB calculations provide no signature
for the erosion of the $Z=40$ magicity.
The present axial HF calculations have resolved this problem,
showing that $Z=40$ is not magic in this region
because the axially deformed states lie lower than the spherical state.
Although the pair correlation has not been taken into account,
the lowest energy of the deformed state is even lower
than the spherical HFB energy in most of these nuclei.
Thus, together with the result for $^{32}$Mg in Ref.~\cite{ref:SNM16},
the validity of M3Y-P6 in predicting magic numbers
is reinforced by the axial HF calculations.
It is remarked that deformation at $^{80}$Zr is reproduced at the MF level,
which has been difficult in the self-consistent MF calculations so far
despite a few exceptions.
This deformation can be regarded as an indirect effect of the tensor force,
as already discussed.

The tensor force acts strongly on the unique-parity orbit,
because of its large $\ell$~\cite{ref:Vtn}.
Moreover, the difference in parity makes the unique-parity orbit almost pure,
with little admixture of other spherical s.p. orbitals.
Therefore the tensor-force effects are more conspicuous
on the unique-parity orbit than on others.
The repulsive tensor force effect becomes stronger
as the unique-parity orbit is occupied.
We have pointed out that, in the lower part of the major shell,
the energy of the deformed state goes up
by this interplay of the tensor force and the unique-parity orbit,
and deformation tends to be delayed as seen in $^{96}$Zr.

In the argument of the so-called `type-II shell evolution'~\cite{ref:type2},
it was indicated that deformation could be driven
by the interplay of the unique-parity orbit and the tensor force,
in addition to the central force.
An excited state of $^{68}$Ni was raised as an example,
in which protons are excited from the $jj$-closed $Z=28$ core
at the deformed states.
The tensor force makes excitation easier.
In the present case, deformation requires excitation of protons
from the $\ell s$-closed $Z=40$ core to $0g_{9/2}$.
As in the difference between $N=20$ and $28$~\cite{ref:SNM16},
the tensor-force effects on deformation are opposite
between the $\ell s$-closed nuclei and the $jj$-closed nuclei.
In Ref.~\cite{ref:QPT}, the type-II shell evolution was extensively discussed
for deformation around $^{100}$Zr.
It could be considered
that the tensor-force effects involving the unique-parity orbit
shown in this article have disclosed a specific effect
of the type-II shell evolution discussed in Ref.~\cite{ref:QPT}.

As mentioned in Sec.~\ref{sec:intro},
the exotic shape with the tetrahedral symmetry was argued
for several Zr nuclei, $^{80}$Zr, $^{96}$Zr
and $^{108,110,112}$Zr~\cite{ref:tetra-Zr96,ref:tetra-Zr110,ref:ZLZZ17}.
Some calculations predicted it take place at the g.s.,
which have been supported by no experimental data.
Although the present calculations do not handle the tetrahedral configuration
explicitly,
we shall give comments on this issue.
At $^{96}$Zr, the $N=56$ magicity as well as the $Z=40$ magicity
are propped up by the tensor force,
which prevent the g.s. deformation including the tetrahedral one.
Furthermore, the tetrahedral shape
implies admixture of the octupole deformation,
which could not be driven without occupation of the unique-parity orbit.
The excitation to the unique-parity orbit gives rise to
loss of the binding energy as an effect of the tensor force.
Therefore, the tensor force will not favor the tetrahedral shape
in the Zr nuclei.

In this article we have constrained ourselves to $N\leq 82$.
Beyond $N=82$, the $n0i_{13/2}$ orbit may enter,
and a bigger s.p. space will be desired
which includes the $\ell=10$ basis-functions.
An interesting subject in $N\geq 82$ is the giant halo
predicted in Ref.~\cite{ref:g-halo}.
We leave it for a future study,
which should properly take account of pair correlations
and coupling to the continuum~\cite{ref:NT18}.
However, it is noted that the neutron drip line for Zr is located at $N=86$,
according to the spherical HFB calculation with M3Y-P6~\cite{ref:NS14}.
This implies that, in the prediction with the M3Y-P6 interaction,
neutron halos in this region will not be formed
by no more than four neutrons, not being huge.

In summary,
we have investigated the shape evolution of Zr nuclei
and effects of the tensor force on it,
by the axial Hartree-Fock calculations
with the M3Y-P6 semi-realistic interaction.
Deformation at $N\approx 40$ and in $60\lesssim N\lesssim 70$ is reproduced.
The former has not been easy for the self-consistent MF calculations so far.
The latter seems to resolve the discrepancy
in the prediction of magic numbers in Ref.~\cite{ref:NS14}.
The sudden transition from prolate to oblate is predicted at $^{114}$Zr,
although recovery of spherical shape via the pairing,
shape mixing or triaxial deformation is not ruled out.
The transition from oblate to spherical shape is predicted at $^{120}$Zr.
For the tensor-force effects,
we have pointed out significant roles of the unique-parity orbit
in the shape evolution.
These effects could be crucial for the doubly-magic nature of $^{96}$Zr
and for the predicted shape transitions at $^{114}$Zr and $^{120}$Zr.

\begin{acknowledgments}
%
The authors are grateful to T. Inakura
for providing the HF results with SkM$^\ast$.
This work is financially supported in part
by JSPS KAKENHI Grant Number~16K05342.
Some of the numerical calculations have been performed on HITAC SR24000
at Institute of Management and Information Technologies in Chiba University.
\end{acknowledgments}


\end{document}